\documentclass[aps,prb,reprint,twocolumn,showpieces,superscriptaddress,prb]{revtex4-2}
\usepackage{amsmath}
\usepackage{graphicx}
\usepackage{lmodern}
\usepackage{amsmath}
\usepackage{color}
\usepackage{amssymb}
\usepackage{bm}
\usepackage{braket}
\usepackage{mathtools}
\usepackage{afterpage}
\usepackage{soul}
\usepackage[caption=false,captionskip=-25pt, position=bottom]{subfig}
\usepackage{ulem}

\usepackage[dvipsnames]{xcolor}
\usepackage[colorlinks=true]{hyperref}
\hypersetup{
  colorlinks=true,
  linkcolor=blue,
  citecolor=blue,
  urlcolor=blue
} 
\makeatletter
\def\maketitle{
\@author@finish
\title@column\titleblock@produce
\suppressfloats[t]}
\makeatother

\begin{document}
\title{Spontaneous vortex lattice due to orbital magnetization in valley polarized superconductors}
\author{Ammar Jahin}
\affiliation{Theoretical Division, T-4, Los Alamos National Laboratory, Los Alamos, New Mexico 87545, USA}
\author{Shi-Zeng Lin}
\affiliation{Theoretical Division, T-4, Los Alamos National Laboratory, Los Alamos, New Mexico 87545, USA}
\affiliation{Center for Integrated Nanotechnologies (CINT), Los Alamos National Laboratory, Los Alamos, New Mexico 87545, USA}
\date{\today}

\date{\today}

\begin{abstract}
In this work, we study the spontaneous formation of a vortex lattice in two-dimensional valley polarized superconductors due to orbital magnetization. The screening of magnetic field is weak for two-dimension superconductors, {allowing for the magnetic flux associated with vortices to penetrate deep into the superconducting region}. The Zeeman coupling between orbital magnetization and magnetic fields associated with vortices leads to the formation of a vortex lattice, once the vortex self-energy is lower than the Zeeman energy. We study the phase diagram and the vortex lattice configuration, and discuss the consequences of the vortex lattice formation {in various experimental setups}. 

\end{abstract} 
\maketitle

\section{Introduction}
It is generally believed that superconductivity is most favored when pairing occurs between electrons with opposite momentum~\cite{BCS}. In two-dimensional materials with valley structure, pairing is considered to occur between electrons with opposite momenta at opposite valleys. Indeed, most experimental observations of superconductivity in two-dimensional materials are consistent with this picture even when the underlying pairing mechanism remains debated. %~\cite{Wu_2018,Stauber_2019,Khalaf_2021,Cr_pel_2021}.

Therefore, it came as a big surprise when superconductivity arising from valley-polarized normal state was observed experimentally in rhombohedral graphene \cite{Long_2024} and twisted MoTe$_2$ \cite{Xu_Sun_Li_Zheng_Xu2025}. In the normal state, due to the trigonal warping, there does not exist a pair of electrons with opposite momentum (up to a total valley momentum) at the Fermi surface. Furthermore, the normal state exhibits a large anomalous Hall conductivity. According to the conventional picture, this makes the {valley-polarized} normal state a bad starting point for superconductivity. As such, these experiments have motivated the search for new theories of superconductivity in valley-polarized metal~\cite{Chou_2024,Geier_2024,Jahin_2024,Wang_2024,Yang_2024,Dong_2024,Paoletti_2025,Yoon_2025,Christos_2025}, which have been largely overlooked in the past. 

In valley-polarized metals, time reversal symmetry is broken, and an orbital magnetization due to the rotation of the electron wavefunction is allowed. For two-dimensional superconductors, the screening of magnetic fields is weak, and the orbital magnetization can have a large effect on the superconducting state. 
The orbital magnetization couples to the magnetic field through the standard Zeeman coupling, i.e., $\mathcal{H}_\text{Zeeman} = -\bm{M}_\text{orb} \cdot \bm{B}$, where $\bm{M}_\text{orb}$ is the orbital magnetization and $\bm{B}$ is the magnetic field. As we will show, this coupling can lead to the formation of a spontaneous vortex lattice in superconductors even in the absence of an external magnetic field.

In this work, we will study the conditions for spontaneous vortex lattice formation in two-dimensional valley-polarized superconductors. We will discuss the magnetic field-temperature phase diagram as well as the possible vortex lattice configurations. The emergence of the spontaneous vortex lattice has consequences for dissipation in the superconducting state. Another particularly interesting consequence is the emergence of Majorana fermions at the vortex core when the superconducting state is chiral~\cite{Alicea_2012}, which is likely the case in rhombohedral graphene \cite{Long_2024} and twisted MoTe$_2$ \cite{Xu_Sun_Li_Zheng_Xu2025}. We will take rhombohedral graphene as an example, but the picture is generally valid for other two-dimensional valley-polarized superconductors.

\section{Physical picture}
Beside electron spin, there exists an orbital contribution to magnetization, which arises due to the rotation of the electron wave function when time-reversal symmetry is broken. The orbital magnetization is a fundamental property of materials and plays a crucial role in understanding various phenomena, such as the quantum Hall effect and topological insulators~\cite{Thonhauser2005, Resta2010, Vanderbilt2018,Thonhauser2011,Xiao2010}.  
The valley polarized state in two dimensional systems generally exhibits orbital magnetization~\cite{Sharpe_2019,Serlin_2020,Chen_2022,Chen_2020,Liu_2019,Ren_2021}. When the spin-orbit coupling is weak, the spin contribution to the magnetization can be neglected due to the lack of spontaneous symmetry breaking in two dimensional. 
% \SZL{We do not need to have such a detailed estimate in Introduction.}\sout{I think the argument is that for the electrons in the valence bands, the lack of spin-orbit coupling makes the spin contribution small, and for the electrons in the conduction bands, even though spin polarized, their density is so small that that it we can neglect it. A quick estimate of of spin magnetization at the relevant filling $n = 10^{16} \text{ m}^{-2}$ can be given by $ m_{\text{spin}} = M_{\text{spin}} \times d \times A = (\mu_B \times \frac{1}{d} \times 10^{16})( d \times 5.2 \times 10^{-20}) = 10^{-4} \mu_B$ which is two orders of magnitude smaller than the orbital magnitization.} 
However, the orbital magnetization which is geometrically confined to be perpendicular to the plane can exhibit long-range order. In the presence of a magnetic field, there is a Zeeman coupling between the orbital magnetization and the magnetic field. For a weak magnetic field, the orbital magnetization may be expanded in powers of the magnetic field, with the zeroth order term being the orbital magnetization without the magnetic field~\cite{Xiao2010}. For a strong field, one needs to compute the orbital magnetization for the Landau levels induced by the field.

The orbital magnetization couples to a weak magnetic field through the standard Zeeman coupling, i.e.,
\begin{equation}\label{eq2}
\mathcal{H}_\text{Zeeman} = -\bm{M}_\text{orb} \cdot \bm{B},
\end{equation}
where $\bm{M}_\text{orb}$ is the orbital magnetization and $\bm{B}$ is the magnetic field. We expect that this coupling can lead to the formation of a spontaneous vortex lattice in superconductors with large orbital magnetization even in the absence of an external magnetic field.

First, we give a simple argument for when to expect the vortex lattice phase to be favored. We neglect the interaction between vortices, and consider the energy of forming a single vortex in a thin film of thickness $d$. The energy cost of a single vortex is given by \cite{PhysRevB.75.064514}
\begin{equation}
  E_\text{vortex} = \frac{\Phi_0^2}{8\pi^2\Lambda} \ln\frac{\Lambda}{\xi}, 
\end{equation}
with an effective penetration depth for thin film $\Lambda=2\lambda^2/d$. The Zeeman energy gain due to the orbital magnetization is given by $E_{\text{Zeeman}}=-M_{orb}\Phi_0 d$ where we have used $\int dr^2 B=\Phi_0$. A vortex is favored when $E_{\text{vortex}} + E_{\text{Zeeman}} < 0$
\begin{equation}\label{eq3}
  \frac{\Phi_0}{16\pi^2\lambda^2} \ln\frac{\Lambda}{\xi} < M_{orb},
\end{equation}
a condition that only depends on the intrinsic penetration depth $\lambda$. Since $\lambda\propto 1/\sqrt{1-T/T_c}$, Eq. \eqref{eq3} can be satisfied at a temperature close to $T_c$. The induced vortices form a lattice due to the repulsion between them. The details of the vortex configuration will be discussed below. 

Note that $\mathbf{B}=\nabla\times\bm{A}$ couples to the superconducting wave function through minimal coupling. The Zeeman coupling Eq. \eqref{eq2} causes an instability toward the formation of a vortex lattice. A similar mechanism for an emergent gauge field has been discussed in the context of skyrmion formation. \cite{PhysRevB.110.104420,Gonçalves_Lin_2024}

\section{Ginzburg-Landau Theory}
Here we elaborate in more detail on the physical picture using the Ginzburg-Landau theory. The Ginzburg-Landau free energy of the system is $\mathcal{F}=\mathcal{F}_{S}+\mathcal{F}_{M}+\mathcal{F}_{\text{orb}}$,
where $\mathcal F_S$ is the free energy associated with the superconducting order parameter
\begin{equation}
  \frac{\mathcal{F}_S}{d} =\int d^2\bm r \left[ \frac{\alpha}{2}  |\Delta|^2 + \frac{\beta}{4} |\Delta|^4 - \frac{1}{2 m} |(i\hbar \bm{\nabla}+\frac{2e}{c}\mathbf{A})\Delta|^2  \right]
\end{equation}
 $\mathcal F_{M}$ include the magnetic coupling and magnetic energy free energy
\begin{equation}
  \frac{\mathcal{F}_M}{d} = -\int \frac{d^2\bm r}{4\pi} \left(  H_a+4\pi M_{\text{orb}} \right) B_z + \int \frac{d^3 \bm r}{8\pi}  |\bm{\nabla}\times\bm{A}|^2
\end{equation}
and $\mathcal{F}_{\text{orb}}(M_{\text{orb}})$ is the free energy of the system associated with $M_{\text{orb}}$ and its gradients. 
Here $\alpha$ is the Ginzburg-Landau parameter, $\Delta$ is the superconducting order parameter, $\bm{A}$ is the vector potential and $\mathbf{B}=\nabla\times\bm{A}$. For the magnetic energy, we need to integrate over the whole space (not just over the sample). Both the orbital magnetization $M_{\mathrm{orb}}$ and the applied field $H_a$ are perpendicular to the film. The pairing symmetry is not important for this discussion, and here we take $\Delta$ to describe a $p+ip$ superconductor. 

For thin film geometry, the effective London penetration depth is $\Lambda=2\lambda^2/d$, which is much larger than the superconducting coherence length $\xi$. When the vortex separation is much larger than $\xi$, or away from $H_{c2}$, we can neglect the variation in the amplitude of $\Delta$. Then $\mathcal{F}_S$ reduces to 
\begin{equation}
  \frac{\mathcal{F}_s}{d} = \int d^2r \left[ \frac{\alpha }{2} |\Delta|^2 + \frac{\beta}{4} |\Delta|^4 + \frac{|\Delta|^2}{2 m} |(\hbar \bm{\nabla}\varphi-\frac{2e}{c}\mathbf{A})|^2  \right]
\end{equation}
where $\varphi$ is the phase of the order parameter $\Delta$.  The part involves $\mathbf{A}$ is the London free energy, and the total free energy can by simplified to \cite{PhysRevB.86.180506}
\begin{align}
  \mathcal{F}_L = \  & d\int d^2r \left[  \frac{1}{8\pi \lambda^2} |(\frac{\Phi_0}{2\pi} \bm{\nabla}\varphi-\mathbf{A})|^2  -  \frac{H_a+4\pi M_{orb}}{4\pi} B_z  \right]\nonumber \\ 
  + & \frac{1}{8\pi}\int dr^3  |\bm{\nabla}\times\bm{A}|^2+\mathcal{F}_{\text{orb}}.
\end{align}
When Eq. \eqref{eq3} is satisfied, the system favors a vortex lattice. However, a vortex lattice with the same polarization is not energetically favored. This is because of the net magnetic energy cost outside the sample. The system can lower the energy by forming domains of vortex lattice with opposite polarization by flipping $M_{\text{orb}}$ or valley polarization at the cost of domain wall energy of $M_{\text{orb}}$ determined by $\mathcal{F}_{\text{orb}}$.

To find the optimal vortex configuration, one needs to minimize energy consisting of single vortex energy, vortex-vortex interaction energy, interaction between vortices and magnetic field generated by the orbital magnetization domain wall, self-interaction energy of the domain wall, and domain wall energy. This has been done in the context of thin magnet-superconductor hybrids \cite{PhysRevLett.88.017001} 
% \sout{In the context of hybrids, the ferromagnetic layer with constant magnetization has no magnetic field outside that layer. This means that the only way the two layers interact is through the Zeeman term. In our case, the magnetic layer is also the superconducting layer, so to speak.  Would the analysis change since now the magnetic field generated by the non-zero magnetization can interact with the superconducting order parameter? } 
at zero applied field $H_a=0$. For large film $L\gg \Lambda$ with $L$ the linear dimension of the system, it is found that the stripe configuration of alternating vortex lattice with opposite polarization has the lowest energy. The optimal width of the stripe, $L_s$, and the corresponding energy are
\begin{equation}
  L_{s}=\frac{\Lambda}{4}\exp\left(\frac{\epsilon_{DW}}{9\tilde{m}^2}+0.42\right),
\end{equation}
\begin{equation}
  E_{s}=-\frac{36\tilde{m}^2 A}{\Lambda}\exp\left(\frac{-\epsilon_{DW}}{9\tilde{m}^2}-0.42\right),
\end{equation}
where $\tilde{m}=M_{\text{orb}}-2E_{\mathrm{vortex}}/3>0$ and $\epsilon_{DW}$ is the linear tension of the orbital magnetization domain wall and $A$ is the area of the sample. 

\begin{figure}
  \includegraphics[scale = 0.3]{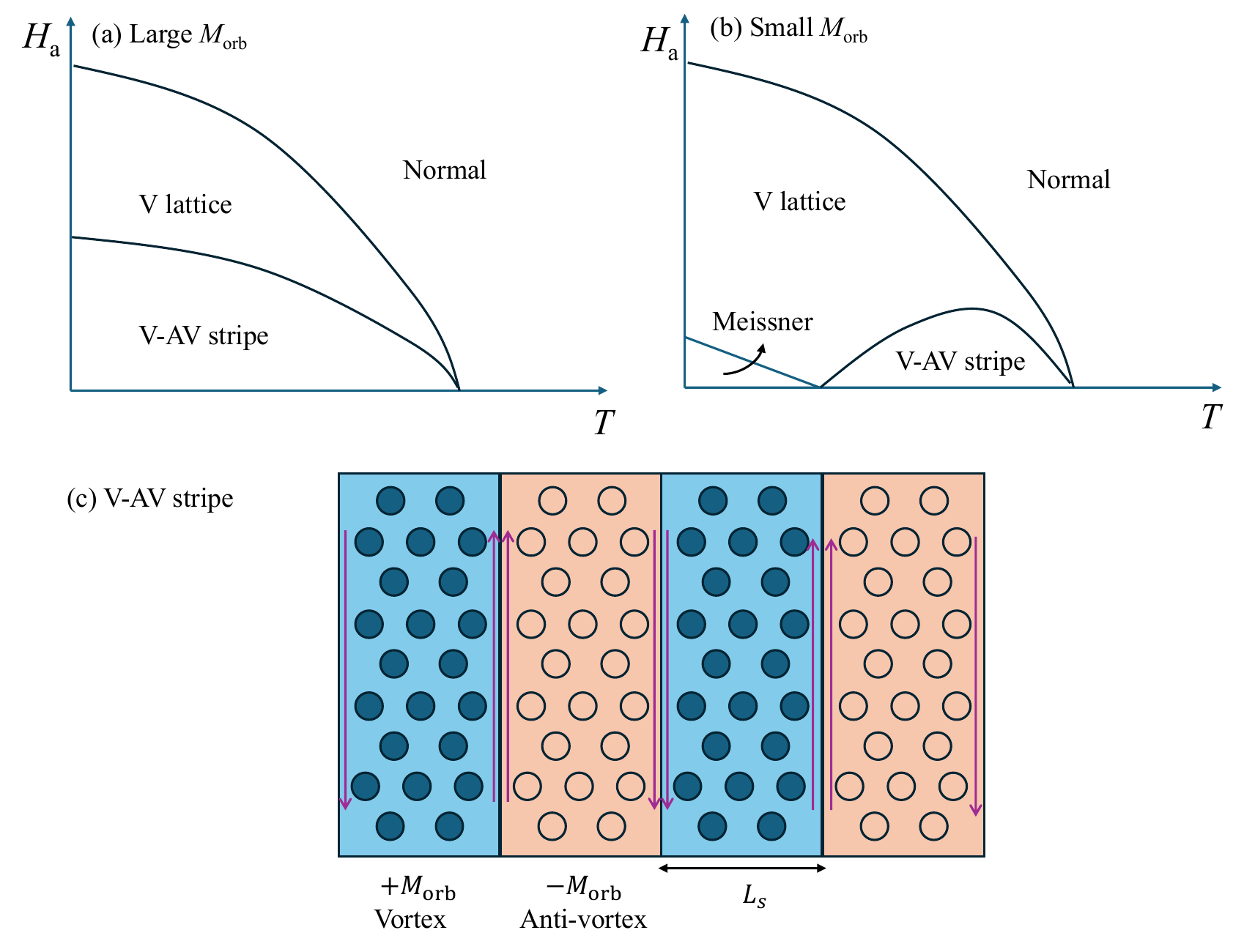}
  \caption{Schematic view of phase diagram when (a) $M_{\mathrm{orb}}$ is large such that spontaneous vortex appears at zero temperature, (b) $M_{\mathrm{orb}}$ is small such that spontaneous vortex only appears near $T_c$. (c) sketches the vortex (V: filled circles) and anti-vortex (AV: open circles) stripe phase where $M_{\mathrm{orb}}$ changes sign between domains. The arrows denote chiral edge modes when the superconducting order parameter is chiral.
  }
  \label{fig1}
  \end{figure}

For a strong $M_{\mathrm{orb}}$, the system develops spontaneously vortex-antivortex stripe at $H_a=0$ and $T=0$. When $M_{\mathrm{orb}}$ is weak, vortex-antivortex stripe only appears near $T_c$ according to Eq. \eqref{eq3}. At zero magnetic field, the stripe width of the opposite vortex polarization region is equal. Across the domain wall, there exists chiral edge mode {due to the distinct topology of the superconducting domains and/or different magnetization orentation across domains}, which is schematically shown in Fig. \ref{fig1} (c). As the magnetic field is increased, the stripe width for the unfavored vortex polarization shrinks, and eventually, the whole system is populated by the favored vortex polarization. Figures \ref{fig1} (a) and (b) illustrate schematically the $H_a$-$T$ phase diagram in the thermodynamic limit. Near $H_{c2}$, the vortex core starts to overlap, and the London approximation employed here is no longer applicable. Near $H_{c2}$, $M_{\text{orb}}$ is just the normal state orbital magnetization. So we have
\begin{equation}
  H_{c2}+4\pi M_{orb}=\Phi_0/(\pi\xi^2).
\end{equation}

When $\Lambda < L<L_s$, we have vortex lattice with the same polarization across the sample. The vortex form a triangular lattice cluster, and we can estimate the vortex separation by optimizing the vortex-vortex interaction and vortex self energy. The energy per area is
\begin{align}
  E_{vc}= \frac{2}{\sqrt{3}r^2}\left( \frac{\Phi_0^2}{8\pi^2\Lambda} \ln\frac{\Lambda}{\xi}-M_{orb}\Phi_0 d  \right.  \qquad \qquad \quad \nonumber \\ 
    +\left.  3\frac{\Phi_0^2}{8\pi^2\Lambda} (H_0(r/\Lambda)-Y_0(r/\Lambda))\right),
\end{align}
where the third term is the vortex-vortex interaction at a distance $r$ with $H_0$ and $Y_0$ the Struve and Bessel functions~\cite{PhysRevB.75.064514}. Optimizing $E_{vc}$ leads to an optimal separation between vortices.

For an estimate, we take $m=m_e$, $d=1$ nm, superfluid density $0.5\times10^{12}/\text{cm}^2$ at $T=0$ K. Using $\lambda=\sqrt{\frac{m}{\mu_0 n_s e^2}}$, we obtain $\lambda\sim 1\ \mathrm{\mu m}$ and $\Lambda\sim 1$ mm. Perpendicular upper critical field for SC1 in experiment is $H_{c2}=0.6 T$, and the coherence length is about $\xi\sim 20$ nm. \cite{Long_2024} When $L< \Lambda$, which is likely the case in the current experiment with current device size is of the order of $1\ \mathrm{\mu m}$, the above estimate is not reliable. The vortex cluster with same polarization is still expected, and its configuration is influenced by the device geometry.

Here we have neglected the vortex core energy, which is valid for thin film since $\Lambda\gg \xi$. 
An improvement to the current estimate is to include the modification of $M_{\text{orb}}$ by the vortex lattice formation, since $M_{\text{orb}}$ depends on superconducting order, see Sec.~\ref{sec:orb_mag_super}. 
However, near $T_c$ or $H_{c2}$, $M_{\text{orb}}$ is just the normal state orbital magnetization.

\begin{figure*}
    \centering
    \captionsetup[subfigure]{oneside,margin={-3.9cm,0cm}, captionskip=-80pt}
    \subfloat[]{\includegraphics[scale=0.55]{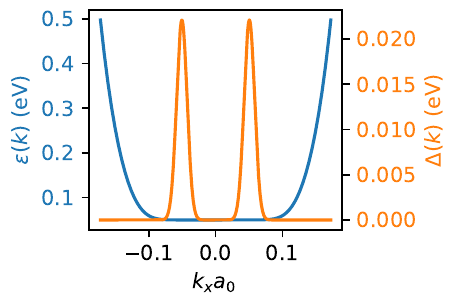}} \hfill
    \captionsetup[subfigure]{oneside,margin={-2.8cm,0cm}, captionskip=-80pt}
    \subfloat[]{\includegraphics[scale=0.55]{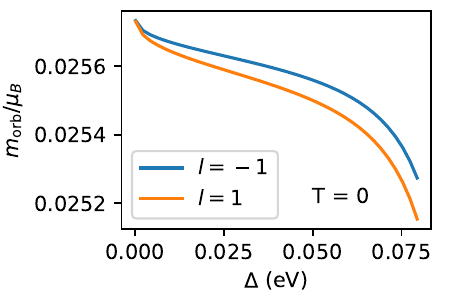}} \hfill
    \captionsetup[subfigure]{oneside,margin={-3.8cm,0cm}, captionskip=-80pt}
    \subfloat[]{\includegraphics[scale=0.55]{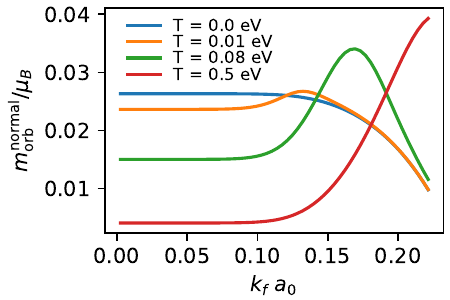}} \hfill
    \captionsetup[subfigure]{oneside,margin={-4.0cm,0cm}, captionskip=-80pt}
    \subfloat[]{\includegraphics[scale=0.55]{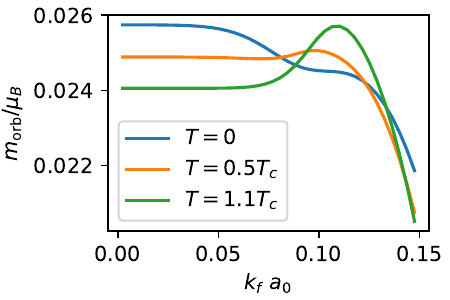}}
    \caption{Orbital magnetization in the normal and superconducting states using the Hamiltonian in Eq.~\eqref{eq:two_band_Hamiltonian} and gap function in Eq.~\eqref{eq:delta_p}.  Panel (a) shows the normal state dispersion and the order parameter. In (b) we plot the magnetization per unit cell of graphene, $m_{\text{orb}} = M_{\text{orb}} A_{uc}$, as a function of $\Delta = \Delta(k = k_f)$ for the angular momenta $l=-1,1$. The evolution of magnatization with temperature is shown in (c). Critical temperature is taken to be $T_c = 0.1 \mu$, and $\Delta = 1.76 T_c$. Here we use $n=5$, $D = 0.05$ eV,  $k_0 = 0.0458$ nm$^{-1}$, and $a_0 = 0.244$ nm is the lattice constant of graphene.} 
    \label{fig:mag_orb_p}
\end{figure*}

\section{Orbital magnetization in the superconducting rhombohedral graphene multilayer} \label{sec:orb_mag_super}

{The key quantity in our physics picture is the orbital magnetization in the superconducting state. As a concrete example, }here we calculate the orbital magnetization in superconducting rhombohedral graphene multilayer, and study how the superconducting order parameters affects the orbital magnetization. We use a two-band model Hamiltonian for $n$-layer rhombohedral graphene~\cite{Herzog_2024},
\begin{align}\label{eq:two_band_Hamiltonian}
    \mathcal H_n(\bm k) = \begin{bmatrix}
        D && \frac{v_0^n}{t_1^{n-1}} (k_x + i k_y)^n \\ 
        \frac{v_0^n}{t_1^{n-1}} (k_x - i k_y)^n && -D 
    \end{bmatrix} - \mu,
\end{align}
where $D$ is the displacement field perpendicular to the plane, $v_0 = 0.542$ eV nm is the velocity of the electron in a single layer of graphene, and $t_1 = 0.355$ eV is the nearest neighbor hopping between two layers. 
The basis of this Hamiltonian is the $A$ sublattice of the top layer and the $B$ sublattice of the bottom layer, and their corresponding annihilation operators for electrons are $c_A(\bm k)$ and $c_B(\bm k)$.
While this Hamiltonian fails to capture the out of plane decay of the wavefunctions, and thus can lead to incorrect overlaps between wavefucntions with different momenta~\cite{Jiang_2025,Bernevig_2025}, it capture the correct dispersion as well as the overall Chern number of the bands, and thus should be a good starting point to study the orbital magnetization. 

The superconducting state is described by the BdG Hamiltonian $H =\sum_{\bm k} \Psi^\dagger (\bm k) \mathcal H_{\text{BdG}}(\bm k) \Psi(\bm k)$, where $\Psi^T(\bm k) = [c_A(\bm k), c_B(\bm k), c^\dagger_A(-\bm k), c^\dagger_B(-\bm k)]$ and the BdG Hamiltonian is
\begin{align}
    \mathcal H_{\text{BdG}} =  \begin{bmatrix}
        \mathcal H_n(\bm k) && \Delta(\bm k) \\ 
        \Delta^\dagger(\bm k) && -\mathcal H^T_n(-\bm k)
    \end{bmatrix}.
\end{align}
The BdG Hamiltonian is diagonalized by a Bogoliubov transformation of the form, 
\begin{align}
    \begin{bmatrix}
        \bm c(\bm k) \\ 
        \bm c^*(-\bm k) 
    \end{bmatrix} 
    = 
    \begin{bmatrix}
        u(\bm k) && v^*(-\bm k) \\ 
        v(\bm k) && u^*(-\bm k)
    \end{bmatrix} 
    \begin{bmatrix}
        \bm \gamma(\bm k) \\ 
        \bm \gamma^*(-\bm k) 
    \end{bmatrix} 
\end{align}
where $\bm c^T(\bm k) = [c_A(\bm k), c_B(\bm k)]$, and $\bm \gamma^T(\bm k) = [\gamma_1(\bm k), \gamma_2(\bm k)]$, such that $H = \sum_{\bm k, n} E_n \gamma^\dagger_n \gamma_n$, with $E_n(\bm k) \geq 0$. 

The formula for the out-of-plane component of the orbital magnetization in the superconducting state is given by~\cite{Joshua_2020}, 
\begin{widetext}
\begin{align}
    M_{\text{orb}} = \frac{e}{d\hbar} \ \sum_n \text{Im} \int \frac{d^2\bm k}{(2\pi)^2} \left[\big \langle \frac{\partial v_{n,\bm k}}{\partial k_x} \big | - \mathcal H^*(-\bm k) + E_n(-\bm k)  \big | \frac{\partial v_{n,\bm k}}{\partial k_y}  \big \rangle (1 - f_{n,\bm k}) + \ \big \langle \frac{\partial u_{n,\bm k}}{\partial k_x} \big | \mathcal H(\bm k) + E_n(\bm k)  \big | \frac{\partial u_{n,\bm k}}{\partial k_y}  \big \rangle  f_{n,\bm k} \right]
\end{align}
\end{widetext}
where $\langle \alpha |u_{n,\bm k} \rangle = u_{\alpha, n}(\bm k)$, $\langle \alpha |v_{n,\bm k} \rangle = v_{\alpha, n}(\bm k)$, $\alpha \in \{A,B\}$, and $f_{n,\bm k}$ is the Femri-Dirac distribution for the excitation with energy $E_n(\bm k)$. 
When $\Delta(\bm k) = 0$, the above formula reduces to the normal state's orbital magnetization, {which can be calculated analytically as shown in Appendix \ref{AppA}}. 
We study how the orbital magnetization is changed in the superconducting state. 

In order to estimate the value of the orbital magnetization in the superconducting state we use the following form of the order parameter,
\begin{align}\label{eq:delta_p}
    \Delta_{l}(\bm k, T) = \Delta_l(T) (k_x + \frac{l}{|l|} ik_y)^{|l|} \frac{e^{-(|k| -k_f)^2/k_0^2}}{k_0} \ \frac{\sigma_0 + \sigma_z}{2}
\end{align}
where we use $\Delta_l(T) = \Delta^0_l \sqrt{1 - T/T_c}$, and $k_0$ is chosen to make sure the order parameter goes to zero away from the Fermi surface. 
In Fig.~\ref{fig:mag_orb_p} (a) we show the dispersion of the normal state of pentalayer graphene, Eq.~\eqref{eq:two_band_Hamiltonian} with $n=5$ and $D = 0.05$ eV. 
The dispersion features a flat part region at the bottom of the conduction band, where the relevant low energy physics is expected to take place. 
In the same plot we also show a typical distribution of $\Delta_l(\bm k, 0)$, Eq.~\eqref{eq:delta_p} for $k_f = 0.2$ nm$^{-1}$. 
The parameter $k_0 = 0.0458$ nm$^{-1}$ in this plot. 
Both the normal state Hamiltonian and the pairing potential contribute to the orbital magnetization. 
In Fig.~\ref{fig:mag_orb_p} (b) we plot the orbital magnetic moment $m_{\text{orb}} = M_{\text{orb}}\times (d A_{uc})$, where $A_{uc}$ is the unit cell area of graphene, as a function of the pairing potential at the Fermi surface which we denote by $\Delta = \Delta(k = k_f, T = 0)$.
First, we observe that the change to the orbital magnetization is small, roughly $~1\%$ of the normal state magnetization, even for large values of $\Delta \approx 0.08 $ eV. {This justifies neglecting the dependence of $M_{\mathrm{orb}}$ on $\Delta$ in the previous section.}
Naively, one would expect the $l = -1, 1$ channels to have opposite effect on the orbital magnetization, however we find that this is not the case.
Both channels reduce the magnetization, albeit at slightly different rates.
This highlight how the relationship between the gap function winding and orbital magnetization is not straightforward. 
A similar observation was made in Ref.~\cite{Liu_2024}. 

The valence and conduction bands of the normal state have opposite Chern numbers, and thus at zero temperature, one expect the orbital magnetization to decrease as we dope electrons into the conduction band. 
The value of the orbital magnetic moment per unit cell is {$m_{\text{orb}} \approx 0.025 \mu_B$, where $\mu_B$ is the Bohr magneton}. 
Interestingly, we find that the orbital magnetization is almost a constant as long as the chemical potential resides in the flat part of the dispersion as shown in Fig.~\ref{fig:mag_orb_p} (c). 
To get an idea of how large this the value is we can compare it to the typical value of magnetization in a ferromagnatic material $M_{\text{typ}} = 100 $ Oe~\cite{Lyuksyutov_2005}. Taking the thickness of the two-dimensional sample to be $1$ nm, we have a magnetization value of approximately $60$ Oe.  

As the temperature increases, the orbital magnetization in the normal state decreases for fillings that are smaller than the temperature energy scale. In Fig.~\ref{fig:mag_orb_p} (c) we see a peak in the orbital magnetization when the chemical potential is comparable to the temperature energy scale.  
Next, we compute how the orbital magnetization depends on temperature in the superconducting state, where the order parameter changes with temperature.
To study this effect, we take $T_c= 0.1 \mu$, and use the BCS relationship $\Delta = 1.76 T_c$ to obtain the superconducting gap at zero temperature. 
The choice of $T_c= 0.1 \mu$ is optimistic, but has been shown to be feasible in the valley polarized superconductors where the quantum geometry can massively enhance superconductivity~\cite{Jahin_2024}. 
Nevertheless, even with such optimistic value of the gap, as shown in Fig.~\ref{fig:mag_orb_p} (d), when the Fermi surface is inside the flat part of the dispersion, the magnetization has a weak dependence on temperature all the way to $T_c$.

\section{Discussion}
We discuss the experimental results \cite{Long_2024} in light of the spontaneous vortex lattice. It is observed that in certain region in the phase diagram, the resistivity in the superconducting state is finite down to the lowest temperature. 
This could be due to the dissipation caused by unpinned vortices driven by an electric current. 
It is also shown that the residue resistivity in the superconducting state fluctuate with time, and an applied magnetic field suppresses the resistivity fluctuations. 
In the vortex-anti-vortex stripe phase, the fluctuations of the orbital magnetization domains cause motion of the domain wall, The vortices follow this motion and induces dissipation. 
For a small device a single domain of orbital magnetization is expected at $T=0$. However, thermal or quantum fluctuations can cause flipping of the orbital magnetization, and changes the polarization of the spontaneous vortex lattice, which also causes dissipation. 
% \sout{Is this a known mechanism? Not sure I can find a reference. Or does this change of magnetization can also be understood in terms of the motion of domain walls? Sounds like a complicated problem. For example, what happens when the votices hit the edge of the sample. } 
The applied magnetic field stabilizes the orbital magnetization, and suppresses the dissipation.

Another important consequence of the vortex lattice is the emergence of Majorana fermion at the vortex cores if the superconductivity is topological~\cite{Alicea_2012} as many theories suggest~\cite{Jahin_2024,Geier_2024,Chou_2024}. These Majorana fermions in the vortex lattice weakly hybridize, and form Majorana fermion band. The presence of vortex lattice can be detected by various experimental techniques, such as scanning tunneling microscopy, scanning SQUID microscopy, and magnetic force microscopy. The anisotropy in the stripe phase can also be detected by transport measurements.

The possibility of spontaneous vortex lattice has been discussed in the context of magnetic superconductors \cite{Bulaevskii_Buzdin_Panjukov_1985} and superconductor-magnet hybrid structures~\cite{Lyuksyutov_Pokrovsky_2005}. In the latter case, the magnetization field generated by localized magnetic moments which are different from the superconducting electrons. As a consequence, there is a direct orbital coupling between the magnetization field and superconducting electrons. In the valley polarized superconductors, the orbital magnetization and the superconductivity are due to the same electrons. The orbital magnetization couples to magnetic field through Zeeman coupling, and the magnetic field orbitally coupled to the superconducting electrons. {Different from the current mechanism}, the spontaneous vortex lattice has also been discussed in graphene multilayer superconductors due to the nontrival superconducting order parameter \cite{PhysRevB.98.195101,PhysRevB.99.195114,Gaggioli_Guerci_Fu_2025}.

\begin{acknowledgements}
  SZL thanks Long Ju for helpful discussions. The work at LANL was carried out under the auspices of the U.S. DOE NNSA under contract No. 89233218CNA000001 through the LDRD Program, and was supported by the Center for Nonlinear Studies at LANL, and was performed, in part, at the Center for Integrated Nanotechnologies, an Office of Science User Facility operated for the U.S. DOE Office of Science, under user proposals $\#2018BU0010$ and $\#2018BU0083$.

\end{acknowledgements}

\appendix

\section{Analytical results for the orbital magnetization in the normal state}\label{AppA}
Here we calculate analytically the orbital magnetization in the normal state. In the normal state, the orbital magnetization is given by
\begin{widetext}
\begin{align}
    M_{\text{orb}} = \frac{e}{2d\hbar} \sum_n \text{Im} \int \frac{d^2 \bm k}{(2\pi)^2} f_n(\bm k) \langle \partial_{\bm k} u_n(\bm k)  | \times \mathcal H(\bm k) + E(\bm k) -2\mu | \partial_{\bm k} u_n(\bm k) \rangle. 
\end{align}
\end{widetext}
The derivatives of $|u_n(\bm k) \rangle$ can be determined using perturpation theory, and are given by
\begin{align}
    | \partial_{\bm k} u_n(\bm k) \rangle = \sum_{m \neq n} |u_n(\bm k) \rangle \frac{\langle u_m(\bm k) | \partial_{\bm k} \mathcal H(\bm k) | u_n(\bm k) \rangle }{E_m(\bm k) - E_n(\bm k)}
\end{align}
For a two band model this expression takes a simple form. 
We start by writing,
\begin{align}
     | \partial_{\bm k} u_0(\bm k) \rangle = \bm a_{0,1}(\bm k) | u_1(\bm k) \rangle,  \\ 
     | \partial_{\bm k} u_1(\bm k) \rangle = \bm a_{1,0}(\bm k) | u_0(\bm k) \rangle,
\end{align}
with
\begin{align}
    \bm a_{m,n} = \frac{\langle u_m(\bm k) | \partial_{\bm k} \mathcal H(\bm k) | u_n(\bm k) \rangle }{E_m(\bm k) - E_n(\bm k)}
\end{align}
Using this we can write 
\begin{align}
    M_{\text{orb}} = \frac{e}{2d\hbar} \sum_n \text{Im} \int \frac{d^2 \bm k}{(2\pi)^2}   (E_1(\bm k) + E_0(\bm k) - 2\mu) \nonumber \\ \times  [ f_0(\bm k) \  \bm a^*_{0,1} \times \bm a_{0,1}  
     + f_1(\bm k)\  \bm a^*_{1,0} \times \bm a_{1,0}  ]   
\end{align}
Noticing that the Berry curvature is given by 
\begin{align}
    \Omega_n(\bm k) = -\text{Im} \ \bm a^*_{n,m} \times \bm a_{n,m}, \qquad m \neq n,
\end{align}
we arrive at the general formula for the orbital magnetization for a two band model, 
\begin{align}
    M_{\text{orb}} = \frac{-e}{2 d \hbar}  \int \frac{d^2 \bm k}{(2\pi)^2}  \ (E_1(\bm k) + E_0(\bm k) - 2\mu)\nonumber \\   \times \ [ f_0(\bm k) \ \Omega_0(\bm k) + f_1(\bm k)\ \Omega_1(\bm k)  ]   . 
\end{align}

Let us consider the case where the Hamiltonian has a {symmetric spectrum} $E_0(\bm k) = - E_1(\bm k)$, for which the expression simplifies significantly, 
\begin{align}
    M_{\text{orb}} = \frac{e \mu}{d\hbar}  \int \frac{d^2 \bm k}{(2\pi)^2} \left[f_0(\bm k) \Omega_0(\bm k) + f_1(\bm k) \Omega_1(\bm k)  \right]. 
\end{align}
Let us focus on the zero temperature limit, where the maximum orbital magnitization is obtained when the chemical potential is just below the conduction band, 
\begin{align}\label{eq:mag_moment}
    m_{\text{orb}} = \frac{e \Delta A C}{4\pi \hbar} = \frac{m_e \Delta A C}{2\pi \hbar^2} \mu_B = \frac{m_e D A n}{2\pi \hbar^2} \mu_B
\end{align}
where we have used $\Delta = 2D$, $C = n/2$ with $n$ being the number of layers. 
Using $D = 50$ meV we get $m_{\text{orb}} = 0.026 \mu_B$ which is in agreement with our numerical results for the normal state.

\bibliography{references}

\begin{thebibliography}{40}%
\makeatletter
\providecommand \@ifxundefined [1]{%
 \@ifx{#1\undefined}
}%
\providecommand \@ifnum [1]{%
 \ifnum #1\expandafter \@firstoftwo
 \else \expandafter \@secondoftwo
 \fi
}%
\providecommand \@ifx [1]{%
 \ifx #1\expandafter \@firstoftwo
 \else \expandafter \@secondoftwo
 \fi
}%
\providecommand \natexlab [1]{#1}%
\providecommand \enquote  [1]{``#1''}%
\providecommand \bibnamefont  [1]{#1}%
\providecommand \bibfnamefont [1]{#1}%
\providecommand \citenamefont [1]{#1}%
\providecommand \href@noop [0]{\@secondoftwo}%
\providecommand \href [0]{\begingroup \@sanitize@url \@href}%
\providecommand \@href[1]{\@@startlink{#1}\@@href}%
\providecommand \@@href[1]{\endgroup#1\@@endlink}%
\providecommand \@sanitize@url [0]{\catcode `\\12\catcode `\$12\catcode `\&12\catcode `\#12\catcode `\^12\catcode `\_12\catcode `\%12\relax}%
\providecommand \@@startlink[1]{}%
\providecommand \@@endlink[0]{}%
\providecommand \url  [0]{\begingroup\@sanitize@url \@url }%
\providecommand \@url [1]{\endgroup\@href {#1}{\urlprefix }}%
\providecommand \urlprefix  [0]{URL }%
\providecommand \Eprint [0]{\href }%
\providecommand \doibase [0]{https://doi.org/}%
\providecommand \selectlanguage [0]{\@gobble}%
\providecommand \bibinfo  [0]{\@secondoftwo}%
\providecommand \bibfield  [0]{\@secondoftwo}%
\providecommand \translation [1]{[#1]}%
\providecommand \BibitemOpen [0]{}%
\providecommand \bibitemStop [0]{}%
\providecommand \bibitemNoStop [0]{.\EOS\space}%
\providecommand \EOS [0]{\spacefactor3000\relax}%
\providecommand \BibitemShut  [1]{\csname bibitem#1\endcsname}%
\let\auto@bib@innerbib\@empty
%</preamble>
\bibitem [{\citenamefont {Bardeen}\ \emph {et~al.}(1957)\citenamefont {Bardeen}, \citenamefont {Cooper},\ and\ \citenamefont {Schrieffer}}]{BCS}%
  \BibitemOpen
  \bibfield  {author} {\bibinfo {author} {\bibfnamefont {J.}~\bibnamefont {Bardeen}}, \bibinfo {author} {\bibfnamefont {L.~N.}\ \bibnamefont {Cooper}},\ and\ \bibinfo {author} {\bibfnamefont {J.~R.}\ \bibnamefont {Schrieffer}},\ }\bibfield  {title} {\bibinfo {title} {Theory of superconductivity},\ }\href {https://doi.org/10.1103/PhysRev.108.1175} {\bibfield  {journal} {\bibinfo  {journal} {Phys. Rev.}\ }\textbf {\bibinfo {volume} {108}},\ \bibinfo {pages} {1175} (\bibinfo {year} {1957})}\BibitemShut {NoStop}%
\bibitem [{\citenamefont {Han}\ \emph {et~al.}(2024)\citenamefont {Han}, \citenamefont {Lu}, \citenamefont {Yao}, \citenamefont {Shi}, \citenamefont {Yang}, \citenamefont {Seo}, \citenamefont {Ye}, \citenamefont {Wu}, \citenamefont {Zhou}, \citenamefont {Liu}, \citenamefont {Shi}, \citenamefont {Hua}, \citenamefont {Watanabe}, \citenamefont {Taniguchi}, \citenamefont {Xiong}, \citenamefont {Fu},\ and\ \citenamefont {Ju}}]{Long_2024}%
  \BibitemOpen
  \bibfield  {author} {\bibinfo {author} {\bibfnamefont {T.}~\bibnamefont {Han}}, \bibinfo {author} {\bibfnamefont {Z.}~\bibnamefont {Lu}}, \bibinfo {author} {\bibfnamefont {Y.}~\bibnamefont {Yao}}, \bibinfo {author} {\bibfnamefont {L.}~\bibnamefont {Shi}}, \bibinfo {author} {\bibfnamefont {J.}~\bibnamefont {Yang}}, \bibinfo {author} {\bibfnamefont {J.}~\bibnamefont {Seo}}, \bibinfo {author} {\bibfnamefont {S.}~\bibnamefont {Ye}}, \bibinfo {author} {\bibfnamefont {Z.}~\bibnamefont {Wu}}, \bibinfo {author} {\bibfnamefont {M.}~\bibnamefont {Zhou}}, \bibinfo {author} {\bibfnamefont {H.}~\bibnamefont {Liu}}, \bibinfo {author} {\bibfnamefont {G.}~\bibnamefont {Shi}}, \bibinfo {author} {\bibfnamefont {Z.}~\bibnamefont {Hua}}, \bibinfo {author} {\bibfnamefont {K.}~\bibnamefont {Watanabe}}, \bibinfo {author} {\bibfnamefont {T.}~\bibnamefont {Taniguchi}}, \bibinfo {author} {\bibfnamefont {P.}~\bibnamefont {Xiong}}, \bibinfo {author} {\bibfnamefont {L.}~\bibnamefont {Fu}},\ and\ \bibinfo {author} {\bibfnamefont {L.}~\bibnamefont {Ju}},\ }\href {https://arxiv.org/abs/2408.15233} {\bibinfo {title} {Signatures of chiral superconductivity in rhombohedral graphene}} (\bibinfo {year} {2024}),\ \Eprint {https://arxiv.org/abs/2408.15233} {arXiv:2408.15233 [cond-mat.mes-hall]} \BibitemShut {NoStop}%
\bibitem [{\citenamefont {Xu}\ \emph {et~al.}(2025)\citenamefont {Xu}, \citenamefont {Sun}, \citenamefont {Li}, \citenamefont {Zheng}, \citenamefont {Xu}, \citenamefont {Gao}, \citenamefont {Jia}, \citenamefont {Watanabe}, \citenamefont {Taniguchi}, \citenamefont {Tong}, \citenamefont {Lu}, \citenamefont {Jia}, \citenamefont {Shi}, \citenamefont {Jiang}, \citenamefont {Zhang}, \citenamefont {Zhang}, \citenamefont {Lei}, \citenamefont {Liu},\ and\ \citenamefont {Li}}]{Xu_Sun_Li_Zheng_Xu2025}%
  \BibitemOpen
  \bibfield  {author} {\bibinfo {author} {\bibfnamefont {F.}~\bibnamefont {Xu}}, \bibinfo {author} {\bibfnamefont {Z.}~\bibnamefont {Sun}}, \bibinfo {author} {\bibfnamefont {J.}~\bibnamefont {Li}}, \bibinfo {author} {\bibfnamefont {C.}~\bibnamefont {Zheng}}, \bibinfo {author} {\bibfnamefont {C.}~\bibnamefont {Xu}}, \bibinfo {author} {\bibfnamefont {J.}~\bibnamefont {Gao}}, \bibinfo {author} {\bibfnamefont {T.}~\bibnamefont {Jia}}, \bibinfo {author} {\bibfnamefont {K.}~\bibnamefont {Watanabe}}, \bibinfo {author} {\bibfnamefont {T.}~\bibnamefont {Taniguchi}}, \bibinfo {author} {\bibfnamefont {B.}~\bibnamefont {Tong}}, \bibinfo {author} {\bibfnamefont {L.}~\bibnamefont {Lu}}, \bibinfo {author} {\bibfnamefont {J.}~\bibnamefont {Jia}}, \bibinfo {author} {\bibfnamefont {Z.}~\bibnamefont {Shi}}, \bibinfo {author} {\bibfnamefont {S.}~\bibnamefont {Jiang}}, \bibinfo {author} {\bibfnamefont {Y.}~\bibnamefont {Zhang}}, \bibinfo {author} {\bibfnamefont {Y.}~\bibnamefont {Zhang}}, \bibinfo {author} {\bibfnamefont {S.}~\bibnamefont {Lei}}, \bibinfo {author} {\bibfnamefont {X.}~\bibnamefont {Liu}},\ and\ \bibinfo {author} {\bibfnamefont {T.}~\bibnamefont {Li}},\ }\bibfield  {title} {\bibinfo {title} {Signatures of unconventional superconductivity near reentrant and fractional quantum anomalous hall insulators},\ }\bibfield  {journal} {\bibinfo  {journal} {arXiv}\ }\href {https://doi.org/10.48550/arXiv.2504.06972} {10.48550/arXiv.2504.06972} (\bibinfo {year} {2025}),\ \bibinfo {note} {arXiv:2504.06972}\BibitemShut {NoStop}%
\bibitem [{\citenamefont {Chou}\ \emph {et~al.}(2024)\citenamefont {Chou}, \citenamefont {Zhu},\ and\ \citenamefont {Sarma}}]{Chou_2024}%
  \BibitemOpen
  \bibfield  {author} {\bibinfo {author} {\bibfnamefont {Y.-Z.}\ \bibnamefont {Chou}}, \bibinfo {author} {\bibfnamefont {J.}~\bibnamefont {Zhu}},\ and\ \bibinfo {author} {\bibfnamefont {S.~D.}\ \bibnamefont {Sarma}},\ }\href {https://arxiv.org/abs/2409.06701} {\bibinfo {title} {Intravalley spin-polarized superconductivity in rhombohedral tetralayer graphene}} (\bibinfo {year} {2024}),\ \Eprint {https://arxiv.org/abs/2409.06701} {arXiv:2409.06701 [cond-mat.supr-con]} \BibitemShut {NoStop}%
\bibitem [{\citenamefont {Geier}\ \emph {et~al.}(2024)\citenamefont {Geier}, \citenamefont {Davydova},\ and\ \citenamefont {Fu}}]{Geier_2024}%
  \BibitemOpen
  \bibfield  {author} {\bibinfo {author} {\bibfnamefont {M.}~\bibnamefont {Geier}}, \bibinfo {author} {\bibfnamefont {M.}~\bibnamefont {Davydova}},\ and\ \bibinfo {author} {\bibfnamefont {L.}~\bibnamefont {Fu}},\ }\href {https://arxiv.org/abs/2409.13829} {\bibinfo {title} {Chiral and topological superconductivity in isospin polarized multilayer graphene}} (\bibinfo {year} {2024}),\ \Eprint {https://arxiv.org/abs/2409.13829} {arXiv:2409.13829 [cond-mat.supr-con]} \BibitemShut {NoStop}%
\bibitem [{\citenamefont {Jahin}\ and\ \citenamefont {Lin}(2025)}]{Jahin_2024}%
  \BibitemOpen
  \bibfield  {author} {\bibinfo {author} {\bibfnamefont {A.}~\bibnamefont {Jahin}}\ and\ \bibinfo {author} {\bibfnamefont {S.-Z.}\ \bibnamefont {Lin}},\ }\href {https://arxiv.org/abs/2411.09664} {\bibinfo {title} {Enhanced kohn-luttinger topological superconductivity in bands with nontrivial geometry}} (\bibinfo {year} {2025}),\ \Eprint {https://arxiv.org/abs/2411.09664} {arXiv:2411.09664 [cond-mat.supr-con]} \BibitemShut {NoStop}%
\bibitem [{\citenamefont {Wang}\ \emph {et~al.}(2024)\citenamefont {Wang}, \citenamefont {Gao},\ and\ \citenamefont {Yang}}]{Wang_2024}%
  \BibitemOpen
  \bibfield  {author} {\bibinfo {author} {\bibfnamefont {Y.-Q.}\ \bibnamefont {Wang}}, \bibinfo {author} {\bibfnamefont {Z.-Q.}\ \bibnamefont {Gao}},\ and\ \bibinfo {author} {\bibfnamefont {H.}~\bibnamefont {Yang}},\ }\href {https://arxiv.org/abs/2410.05384} {\bibinfo {title} {Chiral superconductivity from parent chern band and its non-abelian generalization}} (\bibinfo {year} {2024}),\ \Eprint {https://arxiv.org/abs/2410.05384} {arXiv:2410.05384 [cond-mat.str-el]} \BibitemShut {NoStop}%
\bibitem [{\citenamefont {Yang}\ and\ \citenamefont {Zhang}(2024)}]{Yang_2024}%
  \BibitemOpen
  \bibfield  {author} {\bibinfo {author} {\bibfnamefont {H.}~\bibnamefont {Yang}}\ and\ \bibinfo {author} {\bibfnamefont {Y.-H.}\ \bibnamefont {Zhang}},\ }\href {https://arxiv.org/abs/2411.02503} {\bibinfo {title} {Topological incommensurate fulde-ferrell-larkin-ovchinnikov superconductor and bogoliubov fermi surface in rhombohedral tetra-layer graphene}} (\bibinfo {year} {2024}),\ \Eprint {https://arxiv.org/abs/2411.02503} {arXiv:2411.02503 [cond-mat.supr-con]} \BibitemShut {NoStop}%
\bibitem [{\citenamefont {Dong}\ \emph {et~al.}(2024)\citenamefont {Dong}, \citenamefont {Étienne Lantagne-Hurtubise},\ and\ \citenamefont {Alicea}}]{Dong_2024}%
  \BibitemOpen
  \bibfield  {author} {\bibinfo {author} {\bibfnamefont {Z.}~\bibnamefont {Dong}}, \bibinfo {author} {\bibnamefont {Étienne Lantagne-Hurtubise}},\ and\ \bibinfo {author} {\bibfnamefont {J.}~\bibnamefont {Alicea}},\ }\href {https://arxiv.org/abs/2406.17036} {\bibinfo {title} {Superconductivity from spin-canting fluctuations in rhombohedral graphene}} (\bibinfo {year} {2024}),\ \Eprint {https://arxiv.org/abs/2406.17036} {arXiv:2406.17036 [cond-mat.supr-con]} \BibitemShut {NoStop}%
\bibitem [{\citenamefont {Paoletti}\ \emph {et~al.}(2025)\citenamefont {Paoletti}, \citenamefont {Guerci}, \citenamefont {Sangiovanni}, \citenamefont {Seifert},\ and\ \citenamefont {König}}]{Paoletti_2025}%
  \BibitemOpen
  \bibfield  {author} {\bibinfo {author} {\bibfnamefont {F.}~\bibnamefont {Paoletti}}, \bibinfo {author} {\bibfnamefont {D.}~\bibnamefont {Guerci}}, \bibinfo {author} {\bibfnamefont {G.}~\bibnamefont {Sangiovanni}}, \bibinfo {author} {\bibfnamefont {U.~F.~P.}\ \bibnamefont {Seifert}},\ and\ \bibinfo {author} {\bibfnamefont {E.~J.}\ \bibnamefont {König}},\ }\href {https://arxiv.org/abs/2504.13166} {\bibinfo {title} {Topologically enabled superconductivity: possible implications for rhombohedral graphene}} (\bibinfo {year} {2025}),\ \Eprint {https://arxiv.org/abs/2504.13166} {arXiv:2504.13166 [cond-mat.str-el]} \BibitemShut {NoStop}%
\bibitem [{\citenamefont {Yoon}\ \emph {et~al.}(2025)\citenamefont {Yoon}, \citenamefont {Xu}, \citenamefont {Barlas},\ and\ \citenamefont {Zhang}}]{Yoon_2025}%
  \BibitemOpen
  \bibfield  {author} {\bibinfo {author} {\bibfnamefont {C.}~\bibnamefont {Yoon}}, \bibinfo {author} {\bibfnamefont {T.}~\bibnamefont {Xu}}, \bibinfo {author} {\bibfnamefont {Y.}~\bibnamefont {Barlas}},\ and\ \bibinfo {author} {\bibfnamefont {F.}~\bibnamefont {Zhang}},\ }\href {https://arxiv.org/abs/2502.17555} {\bibinfo {title} {Quarter metal superconductivity}} (\bibinfo {year} {2025}),\ \Eprint {https://arxiv.org/abs/2502.17555} {arXiv:2502.17555 [cond-mat.mes-hall]} \BibitemShut {NoStop}%
\bibitem [{\citenamefont {Christos}\ \emph {et~al.}(2025)\citenamefont {Christos}, \citenamefont {Bonetti},\ and\ \citenamefont {Scheurer}}]{Christos_2025}%
  \BibitemOpen
  \bibfield  {author} {\bibinfo {author} {\bibfnamefont {M.}~\bibnamefont {Christos}}, \bibinfo {author} {\bibfnamefont {P.~M.}\ \bibnamefont {Bonetti}},\ and\ \bibinfo {author} {\bibfnamefont {M.~S.}\ \bibnamefont {Scheurer}},\ }\href {https://arxiv.org/abs/2503.15471} {\bibinfo {title} {Finite-momentum pairing and superlattice superconductivity in valley-imbalanced rhombohedral graphene}} (\bibinfo {year} {2025}),\ \Eprint {https://arxiv.org/abs/2503.15471} {arXiv:2503.15471 [cond-mat.str-el]} \BibitemShut {NoStop}%
\bibitem [{\citenamefont {Alicea}(2012)}]{Alicea_2012}%
  \BibitemOpen
  \bibfield  {author} {\bibinfo {author} {\bibfnamefont {J.}~\bibnamefont {Alicea}},\ }\bibfield  {title} {\bibinfo {title} {New directions in the pursuit of majorana fermions in solid state systems},\ }\href {https://doi.org/10.1088/0034-4885/75/7/076501} {\bibfield  {journal} {\bibinfo  {journal} {Reports on Progress in Physics}\ }\textbf {\bibinfo {volume} {75}},\ \bibinfo {pages} {076501} (\bibinfo {year} {2012})}\BibitemShut {NoStop}%
\bibitem [{\citenamefont {Thonhauser}\ \emph {et~al.}(2005)\citenamefont {Thonhauser}, \citenamefont {Ceresoli}, \citenamefont {Vanderbilt},\ and\ \citenamefont {Resta}}]{Thonhauser2005}%
  \BibitemOpen
  \bibfield  {author} {\bibinfo {author} {\bibfnamefont {T.}~\bibnamefont {Thonhauser}}, \bibinfo {author} {\bibfnamefont {D.}~\bibnamefont {Ceresoli}}, \bibinfo {author} {\bibfnamefont {D.}~\bibnamefont {Vanderbilt}},\ and\ \bibinfo {author} {\bibfnamefont {R.}~\bibnamefont {Resta}},\ }\bibfield  {title} {\bibinfo {title} {Orbital magnetization in periodic insulators},\ }\href@noop {} {\bibfield  {journal} {\bibinfo  {journal} {Physical Review Letters}\ }\textbf {\bibinfo {volume} {95}},\ \bibinfo {pages} {137205} (\bibinfo {year} {2005})}\BibitemShut {NoStop}%
\bibitem [{\citenamefont {Resta}(2010)}]{Resta2010}%
  \BibitemOpen
  \bibfield  {author} {\bibinfo {author} {\bibfnamefont {R.}~\bibnamefont {Resta}},\ }\bibfield  {title} {\bibinfo {title} {Electrical polarization and orbital magnetization: The modern theories},\ }\href@noop {} {\bibfield  {journal} {\bibinfo  {journal} {Journal of Physics: Condensed Matter}\ }\textbf {\bibinfo {volume} {22}},\ \bibinfo {pages} {123201} (\bibinfo {year} {2010})}\BibitemShut {NoStop}%
\bibitem [{\citenamefont {Vanderbilt}(2018)}]{Vanderbilt2018}%
  \BibitemOpen
  \bibfield  {author} {\bibinfo {author} {\bibfnamefont {D.}~\bibnamefont {Vanderbilt}},\ }\href@noop {} {\bibinfo {title} {Berry phases in electronic structure theory: Electric polarization, orbital magnetization and topological insulators}} (\bibinfo {year} {2018})\BibitemShut {NoStop}%
\bibitem [{\citenamefont {Thonhauser}(2011)}]{Thonhauser2011}%
  \BibitemOpen
  \bibfield  {author} {\bibinfo {author} {\bibfnamefont {T.}~\bibnamefont {Thonhauser}},\ }\bibfield  {title} {\bibinfo {title} {Theory of orbital magnetization in periodic insulators},\ }\href {https://doi.org/10.1142/S0217979211058912} {\bibfield  {journal} {\bibinfo  {journal} {International Journal of Modern Physics B}\ }\textbf {\bibinfo {volume} {25}},\ \bibinfo {pages} {1429} (\bibinfo {year} {2011})}\BibitemShut {NoStop}%
\bibitem [{\citenamefont {Xiao}\ \emph {et~al.}(2010)\citenamefont {Xiao}, \citenamefont {Chang},\ and\ \citenamefont {Niu}}]{Xiao2010}%
  \BibitemOpen
  \bibfield  {author} {\bibinfo {author} {\bibfnamefont {D.}~\bibnamefont {Xiao}}, \bibinfo {author} {\bibfnamefont {M.-C.}\ \bibnamefont {Chang}},\ and\ \bibinfo {author} {\bibfnamefont {Q.}~\bibnamefont {Niu}},\ }\bibfield  {title} {\bibinfo {title} {Berry phase effects on electronic properties},\ }\href {https://doi.org/10.1103/RevModPhys.82.1959} {\bibfield  {journal} {\bibinfo  {journal} {Reviews of Modern Physics}\ }\textbf {\bibinfo {volume} {82}},\ \bibinfo {pages} {1959} (\bibinfo {year} {2010})}\BibitemShut {NoStop}%
\bibitem [{\citenamefont {Sharpe}\ \emph {et~al.}(2019)\citenamefont {Sharpe}, \citenamefont {Fox}, \citenamefont {Barnard}, \citenamefont {Finney}, \citenamefont {Watanabe}, \citenamefont {Taniguchi}, \citenamefont {Kastner},\ and\ \citenamefont {Goldhaber-Gordon}}]{Sharpe_2019}%
  \BibitemOpen
  \bibfield  {author} {\bibinfo {author} {\bibfnamefont {A.~L.}\ \bibnamefont {Sharpe}}, \bibinfo {author} {\bibfnamefont {E.~J.}\ \bibnamefont {Fox}}, \bibinfo {author} {\bibfnamefont {A.~W.}\ \bibnamefont {Barnard}}, \bibinfo {author} {\bibfnamefont {J.}~\bibnamefont {Finney}}, \bibinfo {author} {\bibfnamefont {K.}~\bibnamefont {Watanabe}}, \bibinfo {author} {\bibfnamefont {T.}~\bibnamefont {Taniguchi}}, \bibinfo {author} {\bibfnamefont {M.~A.}\ \bibnamefont {Kastner}},\ and\ \bibinfo {author} {\bibfnamefont {D.}~\bibnamefont {Goldhaber-Gordon}},\ }\bibfield  {title} {\bibinfo {title} {Emergent ferromagnetism near three-quarters filling in twisted bilayer graphene},\ }\href {https://doi.org/10.1126/science.aaw3780} {\bibfield  {journal} {\bibinfo  {journal} {Science}\ }\textbf {\bibinfo {volume} {365}},\ \bibinfo {pages} {605–608} (\bibinfo {year} {2019})}\BibitemShut {NoStop}%
\bibitem [{\citenamefont {Serlin}\ \emph {et~al.}(2020)\citenamefont {Serlin}, \citenamefont {Tschirhart}, \citenamefont {Polshyn}, \citenamefont {Zhang}, \citenamefont {Zhu}, \citenamefont {Watanabe}, \citenamefont {Taniguchi}, \citenamefont {Balents},\ and\ \citenamefont {Young}}]{Serlin_2020}%
  \BibitemOpen
  \bibfield  {author} {\bibinfo {author} {\bibfnamefont {M.}~\bibnamefont {Serlin}}, \bibinfo {author} {\bibfnamefont {C.~L.}\ \bibnamefont {Tschirhart}}, \bibinfo {author} {\bibfnamefont {H.}~\bibnamefont {Polshyn}}, \bibinfo {author} {\bibfnamefont {Y.}~\bibnamefont {Zhang}}, \bibinfo {author} {\bibfnamefont {J.}~\bibnamefont {Zhu}}, \bibinfo {author} {\bibfnamefont {K.}~\bibnamefont {Watanabe}}, \bibinfo {author} {\bibfnamefont {T.}~\bibnamefont {Taniguchi}}, \bibinfo {author} {\bibfnamefont {L.}~\bibnamefont {Balents}},\ and\ \bibinfo {author} {\bibfnamefont {A.~F.}\ \bibnamefont {Young}},\ }\bibfield  {title} {\bibinfo {title} {Intrinsic quantized anomalous hall effect in a moiré heterostructure},\ }\href {https://doi.org/10.1126/science.aay5533} {\bibfield  {journal} {\bibinfo  {journal} {Science}\ }\textbf {\bibinfo {volume} {367}},\ \bibinfo {pages} {900–903} (\bibinfo {year} {2020})}\BibitemShut {NoStop}%
\bibitem [{\citenamefont {Chen}\ \emph {et~al.}(2022)\citenamefont {Chen}, \citenamefont {Sharpe}, \citenamefont {Fox}, \citenamefont {Wang}, \citenamefont {Lyu}, \citenamefont {Jiang}, \citenamefont {Li}, \citenamefont {Watanabe}, \citenamefont {Taniguchi}, \citenamefont {Crommie}, \citenamefont {Kastner}, \citenamefont {Shi}, \citenamefont {Goldhaber-Gordon}, \citenamefont {Zhang},\ and\ \citenamefont {Wang}}]{Chen_2022}%
  \BibitemOpen
  \bibfield  {author} {\bibinfo {author} {\bibfnamefont {G.}~\bibnamefont {Chen}}, \bibinfo {author} {\bibfnamefont {A.~L.}\ \bibnamefont {Sharpe}}, \bibinfo {author} {\bibfnamefont {E.~J.}\ \bibnamefont {Fox}}, \bibinfo {author} {\bibfnamefont {S.}~\bibnamefont {Wang}}, \bibinfo {author} {\bibfnamefont {B.}~\bibnamefont {Lyu}}, \bibinfo {author} {\bibfnamefont {L.}~\bibnamefont {Jiang}}, \bibinfo {author} {\bibfnamefont {H.}~\bibnamefont {Li}}, \bibinfo {author} {\bibfnamefont {K.}~\bibnamefont {Watanabe}}, \bibinfo {author} {\bibfnamefont {T.}~\bibnamefont {Taniguchi}}, \bibinfo {author} {\bibfnamefont {M.~F.}\ \bibnamefont {Crommie}}, \bibinfo {author} {\bibfnamefont {M.~A.}\ \bibnamefont {Kastner}}, \bibinfo {author} {\bibfnamefont {Z.}~\bibnamefont {Shi}}, \bibinfo {author} {\bibfnamefont {D.}~\bibnamefont {Goldhaber-Gordon}}, \bibinfo {author} {\bibfnamefont {Y.}~\bibnamefont {Zhang}},\ and\ \bibinfo {author} {\bibfnamefont {F.}~\bibnamefont {Wang}},\ }\bibfield  {title} {\bibinfo {title} {Tunable orbital ferromagnetism at noninteger filling of a moiré superlattice},\ }\href {https://doi.org/10.1021/acs.nanolett.1c03699} {\bibfield  {journal} {\bibinfo  {journal} {Nano Letters}\ }\textbf {\bibinfo {volume} {22}},\ \bibinfo {pages} {238–245} (\bibinfo {year} {2022})}\BibitemShut {NoStop}%
\bibitem [{\citenamefont {Chen}\ \emph {et~al.}(2020)\citenamefont {Chen}, \citenamefont {Sharpe}, \citenamefont {Fox}, \citenamefont {Zhang}, \citenamefont {Wang}, \citenamefont {Jiang}, \citenamefont {Lyu}, \citenamefont {Li}, \citenamefont {Watanabe}, \citenamefont {Taniguchi}, \citenamefont {Shi}, \citenamefont {Senthil}, \citenamefont {Goldhaber-Gordon}, \citenamefont {Zhang},\ and\ \citenamefont {Wang}}]{Chen_2020}%
  \BibitemOpen
  \bibfield  {author} {\bibinfo {author} {\bibfnamefont {G.}~\bibnamefont {Chen}}, \bibinfo {author} {\bibfnamefont {A.~L.}\ \bibnamefont {Sharpe}}, \bibinfo {author} {\bibfnamefont {E.~J.}\ \bibnamefont {Fox}}, \bibinfo {author} {\bibfnamefont {Y.-H.}\ \bibnamefont {Zhang}}, \bibinfo {author} {\bibfnamefont {S.}~\bibnamefont {Wang}}, \bibinfo {author} {\bibfnamefont {L.}~\bibnamefont {Jiang}}, \bibinfo {author} {\bibfnamefont {B.}~\bibnamefont {Lyu}}, \bibinfo {author} {\bibfnamefont {H.}~\bibnamefont {Li}}, \bibinfo {author} {\bibfnamefont {K.}~\bibnamefont {Watanabe}}, \bibinfo {author} {\bibfnamefont {T.}~\bibnamefont {Taniguchi}}, \bibinfo {author} {\bibfnamefont {Z.}~\bibnamefont {Shi}}, \bibinfo {author} {\bibfnamefont {T.}~\bibnamefont {Senthil}}, \bibinfo {author} {\bibfnamefont {D.}~\bibnamefont {Goldhaber-Gordon}}, \bibinfo {author} {\bibfnamefont {Y.}~\bibnamefont {Zhang}},\ and\ \bibinfo {author} {\bibfnamefont {F.}~\bibnamefont {Wang}},\ }\bibfield  {title} {\bibinfo {title} {Tunable correlated chern insulator and ferromagnetism in a moiré superlattice},\ }\href {https://doi.org/10.1038/s41586-020-2049-7} {\bibfield  {journal} {\bibinfo  {journal} {Nature}\ }\textbf {\bibinfo {volume} {579}},\ \bibinfo {pages} {56–61} (\bibinfo {year} {2020})}\BibitemShut {NoStop}%
\bibitem [{\citenamefont {Liu}\ \emph {et~al.}(2019)\citenamefont {Liu}, \citenamefont {Ma}, \citenamefont {Gao},\ and\ \citenamefont {Dai}}]{Liu_2019}%
  \BibitemOpen
  \bibfield  {author} {\bibinfo {author} {\bibfnamefont {J.}~\bibnamefont {Liu}}, \bibinfo {author} {\bibfnamefont {Z.}~\bibnamefont {Ma}}, \bibinfo {author} {\bibfnamefont {J.}~\bibnamefont {Gao}},\ and\ \bibinfo {author} {\bibfnamefont {X.}~\bibnamefont {Dai}},\ }\bibfield  {title} {\bibinfo {title} {Quantum valley hall effect, orbital magnetism, and anomalous hall effect in twisted multilayer graphene systems},\ }\href {https://doi.org/10.1103/PhysRevX.9.031021} {\bibfield  {journal} {\bibinfo  {journal} {Phys. Rev. X}\ }\textbf {\bibinfo {volume} {9}},\ \bibinfo {pages} {031021} (\bibinfo {year} {2019})}\BibitemShut {NoStop}%
\bibitem [{\citenamefont {Ren}\ \emph {et~al.}(2021)\citenamefont {Ren}, \citenamefont {Jiang}, \citenamefont {Qiao},\ and\ \citenamefont {Sheng}}]{Ren_2021}%
  \BibitemOpen
  \bibfield  {author} {\bibinfo {author} {\bibfnamefont {Y.}~\bibnamefont {Ren}}, \bibinfo {author} {\bibfnamefont {H.-C.}\ \bibnamefont {Jiang}}, \bibinfo {author} {\bibfnamefont {Z.}~\bibnamefont {Qiao}},\ and\ \bibinfo {author} {\bibfnamefont {D.~N.}\ \bibnamefont {Sheng}},\ }\bibfield  {title} {\bibinfo {title} {Orbital chern insulator and quantum phase diagram of a kagome electron system with half-filled flat bands},\ }\href {https://doi.org/10.1103/PhysRevLett.126.117602} {\bibfield  {journal} {\bibinfo  {journal} {Phys. Rev. Lett.}\ }\textbf {\bibinfo {volume} {126}},\ \bibinfo {pages} {117602} (\bibinfo {year} {2021})}\BibitemShut {NoStop}%
\bibitem [{\citenamefont {Kogan}(2007)}]{PhysRevB.75.064514}%
  \BibitemOpen
  \bibfield  {author} {\bibinfo {author} {\bibfnamefont {V.~G.}\ \bibnamefont {Kogan}},\ }\bibfield  {title} {\bibinfo {title} {Interaction of vortices in thin superconducting films and the berezinskii-kosterlitz-thouless transition},\ }\href {https://doi.org/10.1103/PhysRevB.75.064514} {\bibfield  {journal} {\bibinfo  {journal} {Phys. Rev. B}\ }\textbf {\bibinfo {volume} {75}},\ \bibinfo {pages} {064514} (\bibinfo {year} {2007})}\BibitemShut {NoStop}%
\bibitem [{\citenamefont {Dong}\ and\ \citenamefont {Levitov}(2024)}]{PhysRevB.110.104420}%
  \BibitemOpen
  \bibfield  {author} {\bibinfo {author} {\bibfnamefont {Z.}~\bibnamefont {Dong}}\ and\ \bibinfo {author} {\bibfnamefont {L.}~\bibnamefont {Levitov}},\ }\bibfield  {title} {\bibinfo {title} {Chiral stoner magnetism in dirac bands},\ }\href {https://doi.org/10.1103/PhysRevB.110.104420} {\bibfield  {journal} {\bibinfo  {journal} {Phys. Rev. B}\ }\textbf {\bibinfo {volume} {110}},\ \bibinfo {pages} {104420} (\bibinfo {year} {2024})}\BibitemShut {NoStop}%
\bibitem [{\citenamefont {Gonçalves}\ and\ \citenamefont {Lin}(2024)}]{Gonçalves_Lin_2024}%
  \BibitemOpen
  \bibfield  {author} {\bibinfo {author} {\bibfnamefont {M.}~\bibnamefont {Gonçalves}}\ and\ \bibinfo {author} {\bibfnamefont {S.-Z.}\ \bibnamefont {Lin}},\ }\bibfield  {title} {\bibinfo {title} {Doping-induced quantum anomalous hall crystals and topological domain walls},\ }\bibfield  {journal} {\bibinfo  {journal} {arXiv}\ }\href {https://doi.org/10.48550/arXiv.2407.12198} {10.48550/arXiv.2407.12198} (\bibinfo {year} {2024}),\ \bibinfo {note} {arXiv:2407.12198}\BibitemShut {NoStop}%
\bibitem [{\citenamefont {Lin}\ \emph {et~al.}(2012)\citenamefont {Lin}, \citenamefont {Bulaevskii},\ and\ \citenamefont {Batista}}]{PhysRevB.86.180506}%
  \BibitemOpen
  \bibfield  {author} {\bibinfo {author} {\bibfnamefont {S.-Z.}\ \bibnamefont {Lin}}, \bibinfo {author} {\bibfnamefont {L.~N.}\ \bibnamefont {Bulaevskii}},\ and\ \bibinfo {author} {\bibfnamefont {C.~D.}\ \bibnamefont {Batista}},\ }\bibfield  {title} {\bibinfo {title} {Vortex dynamics in ferromagnetic superconductors: Vortex clusters, domain walls, and enhanced viscosity},\ }\href {https://doi.org/10.1103/PhysRevB.86.180506} {\bibfield  {journal} {\bibinfo  {journal} {Phys. Rev. B}\ }\textbf {\bibinfo {volume} {86}},\ \bibinfo {pages} {180506} (\bibinfo {year} {2012})}\BibitemShut {NoStop}%
\bibitem [{\citenamefont {Erdin}\ \emph {et~al.}(2001)\citenamefont {Erdin}, \citenamefont {Lyuksyutov}, \citenamefont {Pokrovsky},\ and\ \citenamefont {Vinokur}}]{PhysRevLett.88.017001}%
  \BibitemOpen
  \bibfield  {author} {\bibinfo {author} {\bibfnamefont {S.}~\bibnamefont {Erdin}}, \bibinfo {author} {\bibfnamefont {I.~F.}\ \bibnamefont {Lyuksyutov}}, \bibinfo {author} {\bibfnamefont {V.~L.}\ \bibnamefont {Pokrovsky}},\ and\ \bibinfo {author} {\bibfnamefont {V.~M.}\ \bibnamefont {Vinokur}},\ }\bibfield  {title} {\bibinfo {title} {Topological textures in a ferromagnet-superconductor bilayer},\ }\href {https://doi.org/10.1103/PhysRevLett.88.017001} {\bibfield  {journal} {\bibinfo  {journal} {Phys. Rev. Lett.}\ }\textbf {\bibinfo {volume} {88}},\ \bibinfo {pages} {017001} (\bibinfo {year} {2001})}\BibitemShut {NoStop}%
\bibitem [{\citenamefont {Herzog-Arbeitman}\ \emph {et~al.}(2024)\citenamefont {Herzog-Arbeitman}, \citenamefont {Wang}, \citenamefont {Liu}, \citenamefont {Tam}, \citenamefont {Qi}, \citenamefont {Jia}, \citenamefont {Efetov}, \citenamefont {Vafek}, \citenamefont {Regnault}, \citenamefont {Weng}, \citenamefont {Wu}, \citenamefont {Bernevig},\ and\ \citenamefont {Yu}}]{Herzog_2024}%
  \BibitemOpen
  \bibfield  {author} {\bibinfo {author} {\bibfnamefont {J.}~\bibnamefont {Herzog-Arbeitman}}, \bibinfo {author} {\bibfnamefont {Y.}~\bibnamefont {Wang}}, \bibinfo {author} {\bibfnamefont {J.}~\bibnamefont {Liu}}, \bibinfo {author} {\bibfnamefont {P.~M.}\ \bibnamefont {Tam}}, \bibinfo {author} {\bibfnamefont {Z.}~\bibnamefont {Qi}}, \bibinfo {author} {\bibfnamefont {Y.}~\bibnamefont {Jia}}, \bibinfo {author} {\bibfnamefont {D.~K.}\ \bibnamefont {Efetov}}, \bibinfo {author} {\bibfnamefont {O.}~\bibnamefont {Vafek}}, \bibinfo {author} {\bibfnamefont {N.}~\bibnamefont {Regnault}}, \bibinfo {author} {\bibfnamefont {H.}~\bibnamefont {Weng}}, \bibinfo {author} {\bibfnamefont {Q.}~\bibnamefont {Wu}}, \bibinfo {author} {\bibfnamefont {B.~A.}\ \bibnamefont {Bernevig}},\ and\ \bibinfo {author} {\bibfnamefont {J.}~\bibnamefont {Yu}},\ }\bibfield  {title} {\bibinfo {title} {Moir\'e fractional chern insulators. ii. first-principles calculations and continuum models of rhombohedral graphene superlattices},\ }\href {https://doi.org/10.1103/PhysRevB.109.205122} {\bibfield  {journal} {\bibinfo  {journal} {Phys. Rev. B}\ }\textbf {\bibinfo {volume} {109}},\ \bibinfo {pages} {205122} (\bibinfo {year} {2024})}\BibitemShut {NoStop}%
\bibitem [{\citenamefont {Jiang}\ \emph {et~al.}(2025)\citenamefont {Jiang}, \citenamefont {Heikkilä},\ and\ \citenamefont {Törmä}}]{Jiang_2025}%
  \BibitemOpen
  \bibfield  {author} {\bibinfo {author} {\bibfnamefont {G.}~\bibnamefont {Jiang}}, \bibinfo {author} {\bibfnamefont {T.}~\bibnamefont {Heikkilä}},\ and\ \bibinfo {author} {\bibfnamefont {P.}~\bibnamefont {Törmä}},\ }\href {https://arxiv.org/abs/2504.03617} {\bibinfo {title} {Quantum geometry of the surface states of rhombohedral graphite and its effects on the surface superconductivity}} (\bibinfo {year} {2025}),\ \Eprint {https://arxiv.org/abs/2504.03617} {arXiv:2504.03617 [cond-mat.mes-hall]} \BibitemShut {NoStop}%
\bibitem [{\citenamefont {Bernevig}\ and\ \citenamefont {Kwan}(2025)}]{Bernevig_2025}%
  \BibitemOpen
  \bibfield  {author} {\bibinfo {author} {\bibfnamefont {B.~A.}\ \bibnamefont {Bernevig}}\ and\ \bibinfo {author} {\bibfnamefont {Y.~H.}\ \bibnamefont {Kwan}},\ }\href {https://arxiv.org/abs/2503.09692} {\bibinfo {title} {"berry trashcan" model of interacting electrons in rhombohedral graphene}} (\bibinfo {year} {2025}),\ \Eprint {https://arxiv.org/abs/2503.09692} {arXiv:2503.09692 [cond-mat.str-el]} \BibitemShut {NoStop}%
\bibitem [{\citenamefont {Robbins}\ \emph {et~al.}(2020)\citenamefont {Robbins}, \citenamefont {Annett},\ and\ \citenamefont {Gradhand}}]{Joshua_2020}%
  \BibitemOpen
  \bibfield  {author} {\bibinfo {author} {\bibfnamefont {J.}~\bibnamefont {Robbins}}, \bibinfo {author} {\bibfnamefont {J.~F.}\ \bibnamefont {Annett}},\ and\ \bibinfo {author} {\bibfnamefont {M.}~\bibnamefont {Gradhand}},\ }\bibfield  {title} {\bibinfo {title} {Theory of the orbital moment in a superconductor},\ }\href {https://doi.org/10.1103/PhysRevB.101.134505} {\bibfield  {journal} {\bibinfo  {journal} {Phys. Rev. B}\ }\textbf {\bibinfo {volume} {101}},\ \bibinfo {pages} {134505} (\bibinfo {year} {2020})}\BibitemShut {NoStop}%
\bibitem [{\citenamefont {Liu}\ \emph {et~al.}(2024)\citenamefont {Liu}, \citenamefont {Chatterjee}, \citenamefont {Scaffidi}, \citenamefont {Berg},\ and\ \citenamefont {Altman}}]{Liu_2024}%
  \BibitemOpen
  \bibfield  {author} {\bibinfo {author} {\bibfnamefont {C.}~\bibnamefont {Liu}}, \bibinfo {author} {\bibfnamefont {S.}~\bibnamefont {Chatterjee}}, \bibinfo {author} {\bibfnamefont {T.}~\bibnamefont {Scaffidi}}, \bibinfo {author} {\bibfnamefont {E.}~\bibnamefont {Berg}},\ and\ \bibinfo {author} {\bibfnamefont {E.}~\bibnamefont {Altman}},\ }\bibfield  {title} {\bibinfo {title} {Magnetization amplification in the interlayer pairing superconductor $4hb\text{\ensuremath{-}}{\mathrm{tas}}_{2}$},\ }\href {https://doi.org/10.1103/PhysRevB.110.024502} {\bibfield  {journal} {\bibinfo  {journal} {Phys. Rev. B}\ }\textbf {\bibinfo {volume} {110}},\ \bibinfo {pages} {024502} (\bibinfo {year} {2024})}\BibitemShut {NoStop}%
\bibitem [{\citenamefont {Lyuksyutov~*}\ and\ \citenamefont {Pokrovsky}(2005)}]{Lyuksyutov_2005}%
  \BibitemOpen
  \bibfield  {author} {\bibinfo {author} {\bibfnamefont {I.~F.}\ \bibnamefont {Lyuksyutov~*}}\ and\ \bibinfo {author} {\bibfnamefont {V.~L.}\ \bibnamefont {Pokrovsky}},\ }\bibfield  {title} {\bibinfo {title} {Ferromagnet–superconductor hybrids},\ }\href {https://doi.org/10.1080/00018730500057536} {\bibfield  {journal} {\bibinfo  {journal} {Advances in Physics}\ }\textbf {\bibinfo {volume} {54}},\ \bibinfo {pages} {67–136} (\bibinfo {year} {2005})}\BibitemShut {NoStop}%
\bibitem [{\citenamefont {Bulaevskii}\ \emph {et~al.}(1985)\citenamefont {Bulaevskii}, \citenamefont {Buzdin}, \citenamefont {Kulić},\ and\ \citenamefont {Panjukov}}]{Bulaevskii_Buzdin_Panjukov_1985}%
  \BibitemOpen
  \bibfield  {author} {\bibinfo {author} {\bibfnamefont {L.}~\bibnamefont {Bulaevskii}}, \bibinfo {author} {\bibfnamefont {A.}~\bibnamefont {Buzdin}}, \bibinfo {author} {\bibfnamefont {M.}~\bibnamefont {Kulić}},\ and\ \bibinfo {author} {\bibfnamefont {S.}~\bibnamefont {Panjukov}},\ }\bibfield  {title} {\bibinfo {title} {Coexistence of superconductivity and magnetism theoretical predictions and experimental results},\ }\href {https://doi.org/10.1080/00018738500101741} {\bibfield  {journal} {\bibinfo  {journal} {Advances in Physics}\ }\textbf {\bibinfo {volume} {34}},\ \bibinfo {pages} {175–261} (\bibinfo {year} {1985})}\BibitemShut {NoStop}%
\bibitem [{\citenamefont {Lyuksyutov}\ and\ \citenamefont {Pokrovsky}(2005)}]{Lyuksyutov_Pokrovsky_2005}%
  \BibitemOpen
  \bibfield  {author} {\bibinfo {author} {\bibfnamefont {I.~F.}\ \bibnamefont {Lyuksyutov}}\ and\ \bibinfo {author} {\bibfnamefont {V.~L.}\ \bibnamefont {Pokrovsky}},\ }\bibfield  {title} {\bibinfo {title} {Ferromagnet–superconductor hybrids},\ }\href {https://doi.org/10.1080/00018730500057536} {\bibfield  {journal} {\bibinfo  {journal} {Advances in Physics}\ }\textbf {\bibinfo {volume} {54}},\ \bibinfo {pages} {67–136} (\bibinfo {year} {2005})}\BibitemShut {NoStop}%
\bibitem [{\citenamefont {Su}\ and\ \citenamefont {Lin}(2018)}]{PhysRevB.98.195101}%
  \BibitemOpen
  \bibfield  {author} {\bibinfo {author} {\bibfnamefont {Y.}~\bibnamefont {Su}}\ and\ \bibinfo {author} {\bibfnamefont {S.-Z.}\ \bibnamefont {Lin}},\ }\bibfield  {title} {\bibinfo {title} {Pairing symmetry and spontaneous vortex-antivortex lattice in superconducting twisted-bilayer graphene: Bogoliubov-de gennes approach},\ }\href {https://doi.org/10.1103/PhysRevB.98.195101} {\bibfield  {journal} {\bibinfo  {journal} {Phys. Rev. B}\ }\textbf {\bibinfo {volume} {98}},\ \bibinfo {pages} {195101} (\bibinfo {year} {2018})}\BibitemShut {NoStop}%
\bibitem [{\citenamefont {Wu}(2019)}]{PhysRevB.99.195114}%
  \BibitemOpen
  \bibfield  {author} {\bibinfo {author} {\bibfnamefont {F.}~\bibnamefont {Wu}},\ }\bibfield  {title} {\bibinfo {title} {Topological chiral superconductivity with spontaneous vortices and supercurrent in twisted bilayer graphene},\ }\href {https://doi.org/10.1103/PhysRevB.99.195114} {\bibfield  {journal} {\bibinfo  {journal} {Phys. Rev. B}\ }\textbf {\bibinfo {volume} {99}},\ \bibinfo {pages} {195114} (\bibinfo {year} {2019})}\BibitemShut {NoStop}%
\bibitem [{\citenamefont {Gaggioli}\ \emph {et~al.}(2025)\citenamefont {Gaggioli}, \citenamefont {Guerci},\ and\ \citenamefont {Fu}}]{Gaggioli_Guerci_Fu_2025}%
  \BibitemOpen
  \bibfield  {author} {\bibinfo {author} {\bibfnamefont {F.}~\bibnamefont {Gaggioli}}, \bibinfo {author} {\bibfnamefont {D.}~\bibnamefont {Guerci}},\ and\ \bibinfo {author} {\bibfnamefont {L.}~\bibnamefont {Fu}},\ }\bibfield  {title} {\bibinfo {title} {Spontaneous vortex-antivortex lattice and majorana fermions in rhombohedral graphene},\ }\bibfield  {journal} {\bibinfo  {journal} {arXiv}\ }\href {https://doi.org/10.48550/arXiv.2503.16384} {10.48550/arXiv.2503.16384} (\bibinfo {year} {2025}),\ \bibinfo {note} {arXiv:2503.16384}\BibitemShut {NoStop}%
\end{thebibliography}%

\end{document}